# A Conceptual Framework to Assess the Effectiveness of Rubric Tool

**Phil Smith, Blooma John, Jayan Kurian**
Centre of Commerce and Management
RMIT University
Ho Chi Minh City, Vietnam
Email: phil.smith@rmit.edu.vn; blooma.john@rmit.edu.vn; jayan.kurian@rmit.edu.vn

## Abstract

Rubrics are being used in a wide variety of disciplines in higher education to evaluate the assessments and provide feedback to students. Rubrics are traditionally implemented as paper-based table to grade assessments and provide feedback. The advancement of technology has integrated the Rubric Tool into Learning Management Systems (LMS) like Blackboard to facilitate online marking. In this conceptual paper, we aim to evaluate the effectiveness of the Rubric Tool. We propose an integrated conceptual framework applying various theories from the extant literature to assess the effectiveness of the Rubric Tool. Furthermore, we use the framework to evaluate the effectiveness of the Blackboard LMS Rubric Tool applied in an undergraduate course for grading assessments under the settings of an Australian International University in Asia.

## 1 Introduction

A rubric is defined as a scoring tool that articulates the expectations for a given task in describing levels of quality (Steven and Levi 2013). A rubric tool in general encompasses three distinctive sections which are task description, evaluation criteria, and awarding scores (Reddy and Andrade 2010; Popham 1997). The task description section is used to define an objective to be accomplished whereas the evaluation criteria describe the extent of proficiency required for a given task. Based on the evaluation criterion selected for each task, the scoring section awards the corresponding score. Students value rubrics due to the description of the evaluation criteria for a task and the ability to receive feedback for future improvement. According to Reddy and Andrade (2010) rubrics served as a guide for students to accomplish their tasks incorporating deep learning strategies. Through rubrics, appropriate feedback can be given at an individual task level as well as at an overall assignment level (Hepplestone et al. 2011). Thus, students value rubrics due to the description of the evaluation criteria for a task and the ability to receive feedback.

Literature indicates that rubrics are widely used in several disciplines in higher education (Reddy and Andrade 2010). Rubrics are traditionally implemented as paper-based table to grade assessments and provide feedback. However, as indicated by Czaplewski (2009), the best paper based rubrics are just not entirely self-explanatory to students to perceive and understand that they have been graded fairly. The advancement of technology has integrated Rubric Tool into Learning Management Systems (LMS) like Blackboard to facilitate online marking. Students perceive that the grading is transparent and fair when they receive their detailed grading using Rubric Tool (Andrade and Du 2005). On the other hand, instructors consider consistency, reliability and efficiency of grading as the major value in using the rubrics tool (Campbell 2005). Furthermore, assessment effectiveness using rubrics on a learning management system has been studied by Atkinson and Lim (2013) in terms of structure, feedback, consistency, fairness, and efficiency. The outcome of the study revealed that students were clear with their assessment tasks, scores received, and how they could improve in future assessments. On the other hand, marking efficiency and student satisfaction coupled with this form of feedback generated value for instructors.

There are wide-spread benefits of grading students work electronically, and most importantly, using a Rubric Tool (Campbell 2005). These benefits include reduction in redundant work such as calculation and recording of scores, as well as giving constructive feedback for each task in addition to the specified rubric's evaluation criterion. Some of the other uses of rubrics in addition to grading (Campbell 2005) include improving course delivery and design (Schneider 2006) and in the evaluation of programmes (Dunbar et al. 2006).

Though several studies (Reddy and Andrade 2010; Atkinson and Lim 2013) have examined rubrics as a tool primarily for evaluation, there has been little research on the theoretical grounding of using rubrics for evaluation and to promote student learning. Thus the objective of this of this study is to examine theories that explain characteristics of assessment effectiveness and propose a framework leading to improved student learning.

## 2 LITERATURE REVIEW

To develop a conceptual framework for assessing the effectiveness of the Rubric Tool, various theories are presented in this section.

### 2.1 Technology Acceptance model (TAM)

The Technology Acceptance Model (TAM) is a prominent information systems theory which succinctly explains how users adapt to a new technology (Davis 1989). The theory has two constructs which are "perceived usefulness" and "perceived ease of use". Perceived usefulness correlates with a user's perception that using the new technology will improve their performance, whereas perceived ease of use correlates with a user's perception that the new technology can be used without much difficulty and with greater ease. The matrix rubrics structure presents succinctly the explicit task that students have to accomplish to achieve the learning requirements (Atkinson and Lim 2013). This concurs with the TAM construct – perceived usefulness. The efficiency in using rubrics in terms of understanding levels of performance and the ability to access rubrics online (Atkinson and Lim 2013) even through mobile devices, concurs with the TAM construct - Perceived ease of use.

### 2.2 Learning Theory

The model of Learning Theory is inclusive of five sub-categories of theory, these being behaviourism, cognitivism, social cultural theory, meta-cognitivism, and social constructivism. (Thurlings et al. 2013). In each case appropriate assessment and feedback depends on the category theory applied (Dreher et al. 2011).

Behaviourism uses punishment and reward to amend behaviour. Rubrics apply these to student's work through the mark that is applied, which expresses approval or disapproval, but also communicates the preferred response required to be rewarded. Through Rubrics, appropriate feedback can be given at an individual task level as well as at an overall assignment level (Hepplestone et al. 2011).

Cognitivism in contrast, relies on building new information on existing knowledge and conceptual understanding. Information seeking behaviour, interpretation and analysis are expected in assessments. In this regard a rubric may define, evaluate and approve through grading and feedback the presence of these assessment criteria, each being indicative of students' cognitive processes within the assessment (Mishra 2002). The rubric also guides the student toward the correct response by illustrating the criteria to be met in an excellent answer.

Social Cultural Theory in learning stresses the development of human abilities. These are identified and developed through dialogue between the learner and teacher. Interactive dialogue is critical to achieve learning under this model. As a consequence, learning becomes a social process rather than an individual process; and is culturally influenced, and dialogue based (Renner et al. 2014; Vygotsky 1978). Feedback from the teacher enables the learner to develop to the next higher stage of learning, a pathway to deeper understanding is thus established. The rubric tends to be a one way communication either before or after assessment is completed, and hence cannot leverage the social aspect of feedback. For team assignments it may reward group participation and dialogue although in only limited way.

Metacognitive knowledge requires the development of understanding about one's own cognition. Broadly this will encompass in addition metacognitive awareness, self-awareness, self-reflection and the ability to in effect self-regulate. The control is thus monitoring and control of standards and behaviour is internalised (Pintrich 2002). The application of a rubric by focusing the student's attention on aspects of the task through applying rewards, may encourage such self assessment. In addition, the clarity with which the rubric communicates the criteria and the expression of these through detailed descriptions of what constitutes a poor or high quality response, may teach the student to self-regulate thereby managing quality in future assignments. Self efficacy in writing skills was lifted through rubric use in research on 4th grade students which delivered longer term effects (Andrade et al. 2010; Reddy and Andrade 2010).

Social constructivism is the learning theory that emphasises mechanisms through which learners build knowledge. The learner is actively engaged in constructing knowledge and understanding by relating new knowledge to what they already know and understand (Gielen et al. 2010; Thurlings et al. 2013). The design of the rubric and assessment in concert will encourage through rewards specific cognitive behaviours. Thus appropriately defined criteria may reward for example, a student's efforts to relate past learning to new learning. The rubric may thus support a constructivist approach.

## 2.3 Justice Theory

Rubrics entail consistent and transparent approach in marking which imply perceptions of fairness among students (Atkinson and Lim 2013). Equity Theory or Justice Theory supports the notion that people seek fairness and justice in allocation decisions, for example, that the reward received compares equally to their effort. By referring to similar others people determine if there has been fair distribution (Adams 1963). The concept of fairness can be also be explained using Justice Theory (Colquitt and Chertkoff 2002) especially in terms of distributive justice where outcome is perceived to be fair and transparent for everyone participating in a task. Rubrics can assist by delivering a level of transparency through the use of the detailed marking guides contained within them. These demonstrate clearly to the student the reason a certain mark was allocated. It may therefore be seen as a fair system in that the same quality of assignment is awarded the same mark across the entire student cohort.

Furthermore, studies in addressing fairness feature validity and reliability of the marking process (Parr et al. 2007). Correctly designed rubrics matching the intended learning outcomes of the assessment, provide validity and reliability that is transparent, fair and, seen to be fair.

## 2.4 Cognitive Load Theory

The theory posits that structure of instructional materials play a significant role in reducing users' cognitive load in terms of intrinsic, extraneous, and germane load (Paas et al. 2010; Tasir and Pin 2012). The rubrics' structure and descriptions for each task ensure performance levels that should be accomplished for each level (Atkinson and Lim 2013). Intrinsic load entails that the details of each task can be easily understood without reference to other external elements. Indeed, Andrade and Du (2005) found students experienced reduced assessment anxiety as a result of knowing clearly what was required. Extraneous load is minimised by providing relevant text to students that clearly describes each task.. Germane load ensures that users are able to accomplish a broader overview of the tasks to plan and prepare for future assessments (Paas et al. 2003). To the extent that rubrics assist in skill development and understanding, the effort required for future assignments is also supported.

## 2.5 Communication Theory

Three levels of communications theory are applied to how a Rubric communicates with students. Communication Theory describes the process of communication distinguishing each element from another. In this way the process of information transfer may be analysed. Shannon et al. (1950) point to technical problems, semantic problems and effectiveness problems (Figure 1). The last two referring to respectively: transmission of desired meaning; and whether the message leads to a change in conduct of the receiver (Fiske and Jenkins 2011).

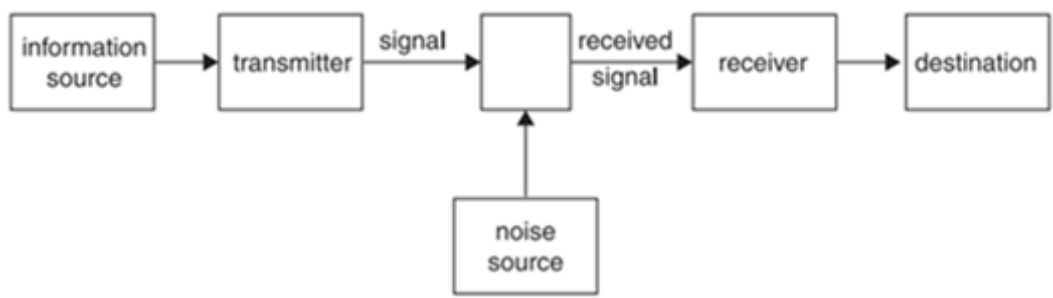

Figure 1: Shannon and Weaver's model of Communication (Adapted from Fiske and Jenkins, 2011)

One technical problem in the process is that of noise, which is anything interfering with a clear signal. In teacher student communication, noise includes poor handwriting, loss or damage to the marked assignment, expression that is abbreviated because of lack of space to write, which may also introduce semantic problems (Shannon et al. 1950). One solution is to use standardised messages as found in rubrics, as a way of combating noise. Also, type-written feedback is always legible. Irrelevant information is similarly regarded as noise in human communication (Edwards 1967).

Thus, based on the literature discussed above, the technology acceptance model can seamlessly explain the efficiency of the rubrics tool in terms of usefulness and ease of use resulting in improved student learning. In terms of providing constructive feedback to students through the Rubrics tool, learning theory is the most appropriate theory to explain improved student learning in the future. Students' perception of fairness in marking can be explained well using the Justice theory which concurs with the transparency of grades awarded through the Rubrics tool in comparison with other students. The structure of the rubrics tool ensures that only a reasonable level of cognitive load, explained by Cognitive Load Theory, is required to perform a task and thus continuously improve student learning.. The consistency in communication especially in terms of feedback leveraged using the rubrics tool ensures that miscommunication is minimized, which can be explained using communication theory. Thus the five theories that describe the effectiveness of the rubrics tool are technology acceptance model, learning theory, justice theory, cognitive load theory, and communication theory. This is illustrated in Figure 2.

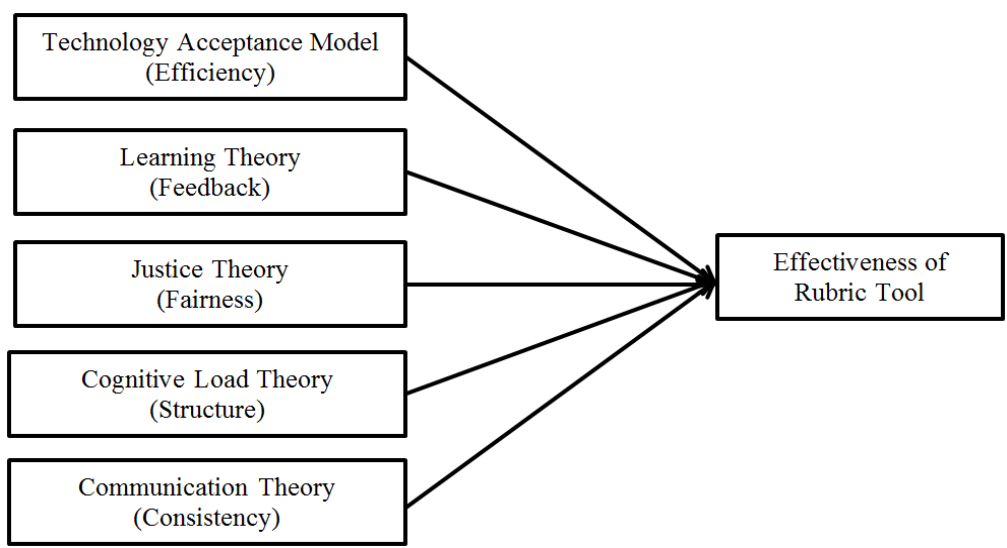

Figure 2: Conceptual Framework to Assess the Effectiveness of Rubric Tool

## 3   DISCUSSION

In this section we present the Rubric Tool used and explained the effectiveness with respect to the features in the framework. The Rubric Tool was applied to a core commerce course offered in an Australian International University located in Asia. The course if offered to more than 300 students by a team of seven faculty members in two different campuses. The team suggested that the tasks of preparing assessments, and marking guides, moderating and marking using a rubric in excel sheet, uploading the marks to grade centre, announcing marks to students, and finally giving feedback to students all included areas of redundancy and reduced efficiency. Thus, to solve these issues with assessments, Blackboard LMS Rubrics Marking Tool was introduced to mark assessments and give feedback to students. The tool was tested and successfully used as a structured, efficient and consistent way of marking. The Rubrics Marking Tool also improved feedback quality by giving fair, timely and constructive assessment feedback to students. A sample rubric used for an assessment in writing a business report is given in Figure 3.

Figure 3: A Sample Rubric Tool used for assessment and feedback

In terms of efficiency, explained by the technology acceptance model, authors were able to grade student's submissions online and record the grade permanently in the system with much ease. The grades were released to all 300 students online and at the same time in a consistent way. In case of amendments, such changes were seamlessly completed using the rubrics tool due to single touch points in the recording of grades. This concurs with the technology acceptance model constructs of perceived ease of use and perceived usefulness. The presentation of rubrics in the form of rows and columns with a grade range extending from unsatisfactory to excellent might be useful for students to evaluate their work and receive feedback for each section with much ease. Faculty was able to give constructive feedback for each task in the assessment in-addition to the evaluation criteria mentioned in the rubrics tool. This helped students to improve their learning and perform well in future assessments which concur with the feedback component of the Learning Theory. In terms of student grading, faculty felt that they have graded assessments without any bias due to the evaluation criteria explained in the rubrics tool which concur with the fairness component of the Justice Theory. From students perspective receiving customized feedback for each task in addition to the general evaluation criteria ensures sufficient justification for the awarded score. The clear structure of the rubrics tool along with the evaluation criteria and grading pattern ensure that variation in grading is minimized with the faculty experiencing less cognitive load in grading. This concurs with the structuring component of the cognitive load theory. The clear structure of the rubrics tool also ensures that students have less cognitive load in understanding the tasks and associated feedback received for each task. Providing constructive feedback is one of the prominent components of improving students' learning. The feedback provided should be easily understood by students without much difficulty in terms of the language used, and should describe how amendments can be made to the existing work to improve grades in future assessments. Thus feedback should be consistent in nature which ensures its value in improving students' learning which concurs with the propositions of communications theory. Though faculty perceptions are explained using the theories described above, the perception of students will be examined only after receiving necessary ethics approval from the human research ethics committee of the university. We intend to accomplish this through a questionnaire survey of the sample population. The survey questions will based on the theoretical constructs described above and illustrated in Figure 2. This will be an avenue for future work.

## 4   Conclusion

This paper proposes a conceptual framework for examining the effectiveness of the rubrics tool in learning. The theoretical grounding for this framework is based on the technology acceptance model, learning theory, justice theory, cognitive load theory, and communication theory. The framework is then used to examine the perception of both faculty and students. Theories used in constructing the framework has been explained in terms of efficiency, feedback, fairness, structure, and consistency which has been adapted from literature on assessment effectiveness. This is a research-in-progress paper and at this stage only faculty perception is discussed.. The student perception will be examined after receiving necessary ethics approval from the human research ethics committee. This study contributes to practitioners in terms of suggestions to improve the design of general e-marking tools to improve student learning. From a theoretical perspective this study adds to the literature on integrating existing information systems theory, learning theory, political philosophy and communication theory to examine the effectiveness of rubrics tool in student learning.

## Copyright